\documentclass[prl,twocolumn,superscriptaddress,showpacs,amscd,amsmath,amssymb,verbatim]{revtex4}

\usepackage{graphicx,color}
\usepackage{epstopdf}
\usepackage{textcomp}
\usepackage[OT1,T1]{fontenc}

\begin{document}
\newcommand{\SGS}{Sm$_3$Ga$_{2.63}$Al$_{2.37}$SiO$_{14}$}
\newcommand{\NGS}{Nd$_3$Ga$_5$SiO$_{14}$}
\newcommand{\PGS}{Pr$_3$Ga$_5$SiO$_{14}$}

\title{Quantum Tunneling in Half-Integer-Spin Kagome-Lattice Langasites}
\author{A. Zorko}
\affiliation{Jo\v{z}ef Stefan Institute, Jamova 39, 1000 Ljubljana, Slovenia}
\affiliation{EN--FIST Centre of Excellence, Dunajska 156, SI-1000 Ljubljana, Slovenia}
\author{F. Bert}
\affiliation{Laboratoire de Physique des Solides, Universit\'e Paris-Sud 11, UMR CNRS 8502, 91405
Orsay, France}
\author{P. Mendels}
\affiliation{Laboratoire de Physique des Solides, Universit\'e Paris-Sud 11, UMR CNRS 8502, 91405 Orsay, France}
\affiliation{Institut Universitaire de France, 103 Boulevard Saint-Michel, F-75005 Paris, France}
\author{A. Poto\v{c}nik}
\affiliation{Jo\v{z}ef Stefan Institute, Jamova 39, 1000 Ljubljana, Slovenia}
\author{A. Amato}
\affiliation{Laboratory for Muon Spin Spectroscopy, Paul Scherrer Institut, CH-5232 Villigen PSI, Switzerland}
\author{C. Baines}
\affiliation{Laboratory for Muon Spin Spectroscopy, Paul Scherrer Institut, CH-5232 Villigen PSI, Switzerland}
\author{K. Marty}
\affiliation{SPINTEC, (UMR8191 CEA/CNRS/UJF/G-INP), CEA Grenoble, INAC, 38054 Grenoble, France}
\author{P. Bordet}
\affiliation{Institut N\'eel, CNRS and Universit\'e Joseph Fourier, BP 166, 38042 Grenoble, France}
\author{P. Lejay}
\affiliation{Institut N\'eel, CNRS and Universit\'e Joseph Fourier, BP 166, 38042 Grenoble, France}
\author{E. Lhotel}
\affiliation{Institut N\'eel, CNRS and Universit\'e Joseph Fourier, BP 166, 38042 Grenoble, France}
\author{V. Simonet}
\affiliation{Institut N\'eel, CNRS and Universit\'e Joseph Fourier, BP 166, 38042 Grenoble, France}
\author{R. Ballou}
\affiliation{Institut N\'eel, CNRS and Universit\'e Joseph Fourier, BP 166, 38042 Grenoble, France}

\date{\today}
\begin{abstract}
Employing the muon spin relaxation technique we evidence temperature independent magnetic fluctuations persisting down to the lowest temperatures in the samarium-based ($J=5/2$) kagome-lattice Langasite. A detailed bulk-magnetization characterization and comparison to the neodymium-based ($J=9/2$) compound allow us to assign the persistent spin dynamics to a quantum tunneling process. This is facilitated by pairwise anisotropic magnetic interactions, leading to a universal scaling of the muon relaxation. Our study reveals a remarkable analogy between weakly interacting half-integer-spin rare-earth magnets and molecular nanomagnets.
\end{abstract}
\maketitle

Quantum tunneling (QT) mechanism allows transitions between different metastable ground states (GSs) through a potential barrier \cite{Hemmen}, which represents a salient alternative to classical thermally activated jumps over the barrier. It has been deeply investigated in molecular nanomagnets (MNMs) containing several magnetic ions entangled into a collective spin state \cite{MN,Sessoli,Thomas}, because of their high application potential in spintronics \cite{Bogani}. Recently, lanthanide compounds have been highlighted as mono-nuclear systems that may magnetically behave like MNMs \cite{Ishikawa,AlDamen}. Similarly to many MNMs they generally possess large magnetic moments and large inherent magnetocrystalline anisotropy (compared to magnetic interactions) setting the potential barrier, which can even be tuned by chemical design to match challenging requirements of quantum computation \cite{Perez}.

\begin{figure}[b]
\includegraphics[trim = 5mm 1mm 0mm 4mm, clip, width=0.9\linewidth]{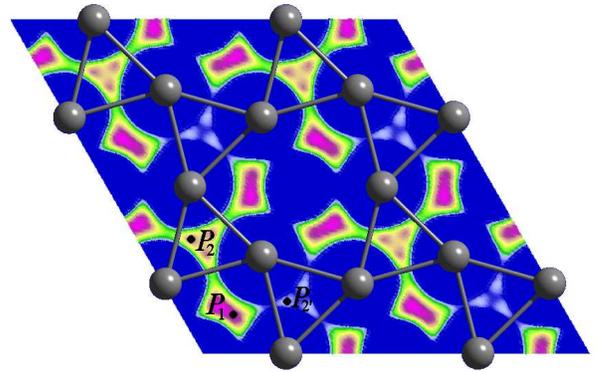}
\caption{Distorted kagome network of rare-earth $R^{3+}$ magnetic moments in Langasites, $R_3BC_3D_2$O$_{14}$ ($ab$ plane at $c=0$). Contour shows the electrostatic potential of La$_3$Ga$_5$SiO$_{14}$ with blue (magenta) regions corresponding to high (low) values. Positions $P_1$ and $P_2$ are most favorable to muons.}
\label{fig1}
\end{figure}

In our study we have focused on the neodymium- and samarium-based members of the Langasite family $R_3BC_3D_2$O$_{14}$ ($R$ -- rare earth) \cite{Bordet}, the first physical realizations of a geometrically frustrated kagome lattice (Fig.~\ref{fig1}) with dominant magnetocrystalline anisotropy.
The Nd-based Langasite \NGS~(NGS) has been intensively studied in recent years \cite{Bordet,Robert,ZhouNd,ZorkoNd,Simonet,Xu,Ghosh,Li} due to its possible spin-liquid GS. Muon spin relaxation ($\mu$SR) \cite{ZorkoNd} alongside neutron scattering \cite{ZhouNd} indeed unambiguously demonstrated the absence of electronic-spin freezing down to at least 60~mK. Since large antiferromagnetic exchange interactions between Nd$^{3+}$ magnetic moments ($J=\frac{9}{2}$) were initially predicted \cite{Bordet}, this was attributed to strong geometrical frustration. Subsequent refined investigations, however, suggested exchange coupling in a sub-Kelvin range \cite{Simonet,Xu}, although the exact value could not be determined.
A plateau in spin relaxation evidencing persistent spin dynamics was detected by $\mu$SR \cite{ZorkoNd} and neutron spin-echo (NSE) experiments \cite{Simonet} below 10~K, i.e., a temperature seemingly unrelated to at least an order of magnitude smaller exchange interaction and an order of magnitude larger crystal-field gap between the GS Kramers doublet and the first excited state \cite{ZhouNd,Simonet}. The fluctuating GS was found to be surprisingly sensitive to the applied magnetic field, as a field-induced crossover from a dynamical to a more static state was observed \cite{ZorkoNd} and even field-induced partial or short-range order was proposed \cite{Ghosh,Li}. Furthermore, zero-field low-temperature heat-capacity data suggested a Schottky anomaly that would require splitting of the GS doublet \cite{Simonet}. Despite all the efforts a comprehensive explanation of these unusual findings is still missing. 

In this Letter, we unveil that the persistent spin dynamics are intrinsic to half-integer-spin Langasites as well as we provide further insight to the unusual magnetic-field dependence of the magnetic fluctuations. We focus on the highly sensitive local-probe $\mu$SR technique \cite{Yaouanc} that has already proven invaluable in Langasites \cite{ZorkoNd,ZorkoPr}. Investigating the new Kramers-ion representative \SGS~(SGS) with Sm$^{3+}$ ($J=\frac{5}{2}$) magnetic moments along with combining new and previously existing \cite{ZorkoNd} $\mu$SR data on NGS enables us to clarify the origin of the persistent spin dynamics. These arise from the quantum tunneling process, leading to a universal scaling of the field-dependent relaxation with the magnitude of the magnetic moments.

In our investigation, the same NGS powder samples were used as previously \cite{ZorkoNd, Simonet}, while the SGS powder samples were synthesized for the first time \cite{sup}. Our syntheses led to highly crystalline samples with the dominant phase \SGS, and minor impurity phases Sm$_3$Ga$_5$O$_{12}$ and Sm$_{4.66}$O(SiO$_4$)$_3$ \cite{sup}. Bulk magnetic properties between 1.6 and 830~K were checked with a commercial Quantum Design MPMS SQUID magnetometer and two purpose-built axial extraction magnetometers. Low-temperature measurements down to 60~mK were performed with a home-built SQUID magnetometer equipped with a miniature dilution refrigerator. 

\begin{figure}[t]
\includegraphics[trim = 1mm 5mm 1mm 7mm, clip, width=1\linewidth]{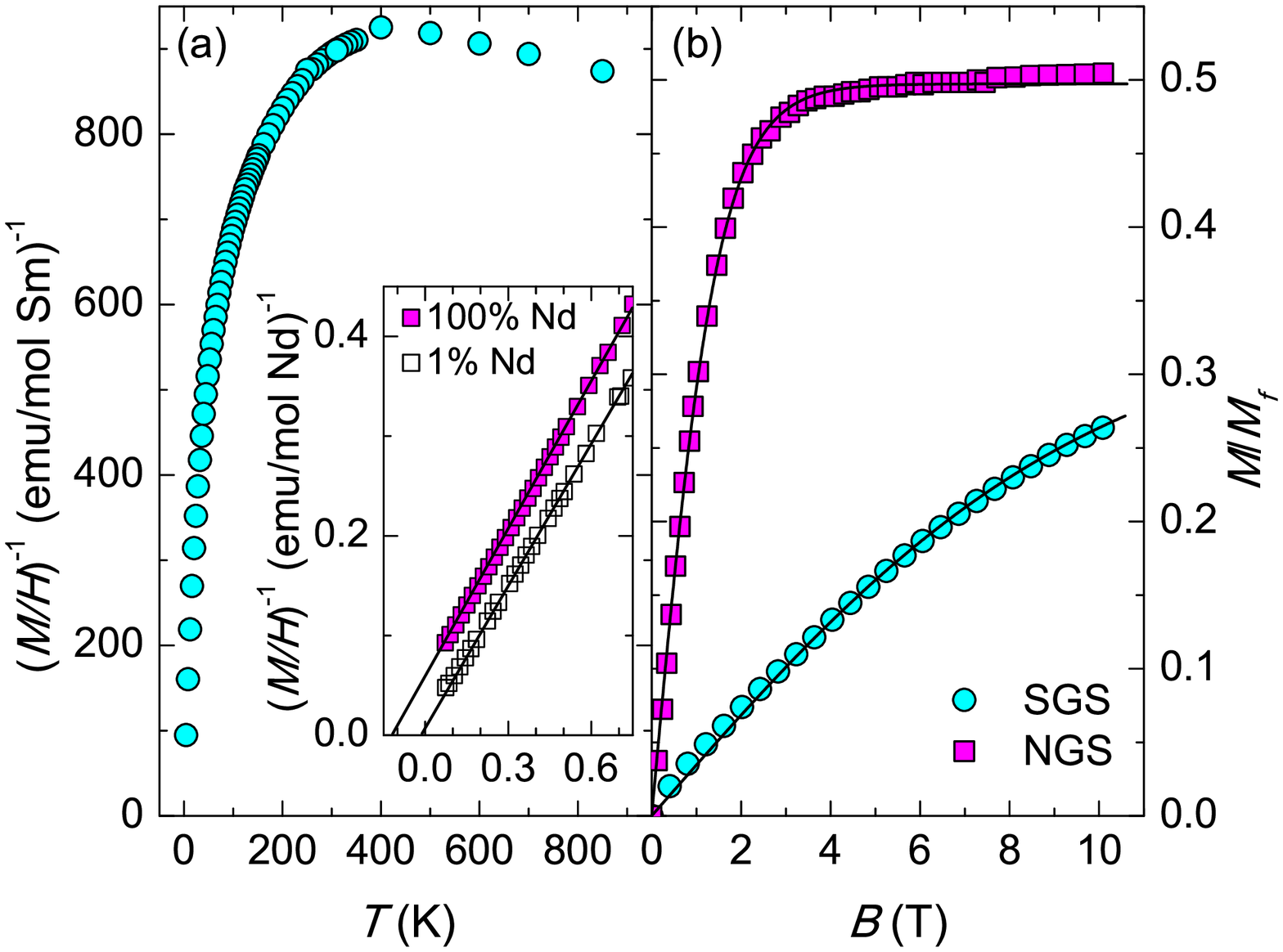}
\caption{(a) Temperature dependence of $(M/H)^{-1}$ (M is magnetization) in \SGS~(SGS) measured in the field  $\mu_0H=1$~T below 350 K and extracted from the $M(H)$ curves above 350 K. Inset shows $(M/H)^{-1}$ measured in 10~mT in \NGS~(NGS) and magnetically diluted (Nd$_{0.01}$La$_{0.99}$)$_{3}$Ga$_{5}$SiO$_{14}$, with the linear extrapolation intersecting the temperature axis. (b) Field dependence of magnetization normalized by the saturated value for free ions measured at 2~K on a powder SGS sample and at 1.6~K on a NGS single crystal ($\bf{H}\Vert\bf{c}$) \cite{Bordet}. Solid lines are fits to the Brillouin function $\frac{\mu_g}{\mu_f}\cdot B_\frac{1}{2}(\mu_g B/k_{\rm B} T)$.}
\label{fig2}
\end{figure}

Like in NGS, bulk susceptibility ($\chi=M/H$ in the linear regime; $M$ is magnetization and $H$ is the applied magnetic field) shows no indications of spin freezing in SGS down to 2~K [Fig.~\ref{fig2}(a)]. A maximum in $(M/H)^{-1}$ observed around 400~K is due to the proximity of the first excited $^6$H$_\frac{7}{2}$ and the GS $^6$H$_\frac{5}{2}$ $J$-multiplet of Sm$^{3+}$\cite{Arajs}. The latter is further split by a low-symmetry local crystal field. Therefore, at low temperatures only the GS Kramers doublet is populated, yielding an effective $\tilde S = \frac{1}{2}$ spin state for both NGS and SGS. This is evidenced by the low-temperature field dependence of magnetization that fits nicely to the normalized Brillouin function $\frac{\mu_g}{\mu_f}\cdot B_\frac{1}{2}(\mu_g B/k_{\rm B} T)$ [Fig.~\ref{fig2}(b)], where $\mu_g$ and $\mu_f=g_J\mu_BJ$ are the GS and the free-ion magnetic moment, $\mu_B$ is the Bohr magneton, $k_B$ the Boltzmann constant and  $B_\frac{1}{2}(x)= \tanh (x)$. The fit yields magnetic moments $\mu_g^{\rm NGS}=1.6\mu_B$ \cite{Bordet,Xu} and $\mu_g^{\rm SGS}=0.25(3)\mu_B$. To obtain the latter we take into account powder averaging and assume analogy to NGS where the moment perpendicular to the $c$ axis is $\mu_\perp^{\rm NGS}=0.8\mu_g^{\rm NGS}$ \cite{Bordet}.
The suitability of the Brillouin function indicates weak magnetic interactions. In NGS, these can be evaluated by comparing the low-temperature susceptibility of the pure and a diluted NGS sample [inset in Fig.~\ref{fig2}(a)], in which non-magnetic La$^{3+}$ ions randomly substitute for 99\% of the Nd$^{3+}$ ions \cite{Simonet}. The Curie-Weiss behavior $(M/H)^{-1}=(T-\theta)/C$ yields the Weiss temperature $\theta=-120(20)$~mK in the pure and $\theta\sim 0$ in the diluted sample.

Invaluable local insight to magnetism in both materials is provided by $\mu$SR. Measurements were performed on the General Purpose Surface-Muon (GPS) and the Low Temperature Facility (LFT) instruments at the Swiss Muon Source (S$\mu$S), Paul Scherrer Institute (PSI), Switzerland. In $\mu$SR experiments the spin of almost 100\% spin-polarized muons is used to locally probe the magnetic field $B_\mu$ at a muon stopping site, causing precession of the muon spin. Static fields lead to coherent oscillations of muon polarization. Their distribution dampens the oscillations, leading to a single dip in the case of random frozen fields \cite{Yaouanc}.  On the other hand, for fast fluctuations of $B_\mu$ ($\nu \gg \gamma_\mu B_\mu$, where $\nu$ is the fluctuation frequency and $\gamma_\mu = 851.6$~MHz/T is the muon gyromagnetic ratio) the muon polarization generally decays exponentially, $P ={\rm e}^{-\lambda t}$, in zero (ZF) or longitudinal magnetic field (LF) $B_{\rm LF}$. The muon spin-lattice relaxation rate
$\lambda=\int_0^\infty \gamma_\mu^2\langle B_{\perp }(t)B_{\perp }(0)\rangle{\rm cos}(\gamma_\mu B_{\rm LF}t){\rm d}t$
measures a spectral-width of spin fluctuations through correlation of the local magnetic field $B_{\perp }$ perpendicular to the applied field.  Description of these correlations with a single correlation time yields the Redfield's relation \cite{Yaouanc}
\begin{equation}
\lambda = \frac{2\gamma_\mu^2B_{\perp }^2\nu}{\nu^2+\gamma_\mu^2 B_{\rm LF}^2}.
\label{eq2}
\end{equation}

\begin{figure}[t]
\includegraphics[trim = 1mm 4mm 1mm 7mm, clip, width=1\linewidth]{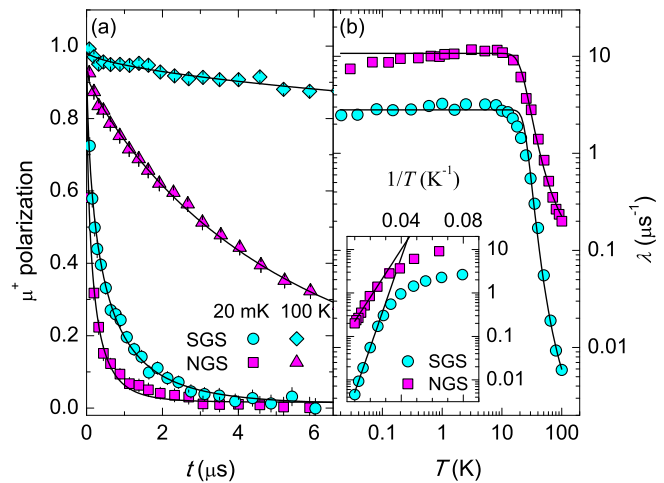}
\caption{(a) Relaxation of ZF muon polarization in \SGS~(SGS) and \NGS~(NGS) \cite{ZorkoNd} at selected temperatures. Lines correspond to the stretched-exponential fit $P(t) ={\rm e}^{-(\lambda t)^\alpha}$. (b) $T$-dependence of the muon spin-lattice relaxation rate with fits (lines) to Eq.~(\ref{eq4}). Inset: the activated behavior $\lambda \propto {\rm e}^{\Delta/T}$ at high temperatures.}
\label{fig3}
\end{figure}

As previously shown for NGS \cite{ZorkoNd}, also for SGS the experimental decay of the muon polarization can be fitted with the stretched-exponential relaxation function
$P(t) ={\rm e}^{-(\lambda t)^\alpha}$
with constant $\alpha=0.6(1)$ in the range between 20~mK and 100~K [Fig.~\ref{fig3}(a)], once the small nonrelaxing background part $\sim$0.05 is subtracted. The stretch exponent $\alpha<1$ is explained by multiple stopping sites in Langasites, as argued later on. The monotonic decay immediately discloses dynamical local fields even at the lowest experimentally accessible temperatures.
$T$-dependence of $\lambda$ measured in ZF reveals strong variations of spin fluctuations at high temperatures in both compounds. The observed activated behavior  $\lambda \propto 1/\nu \propto {\rm e}^{\Delta/T}$ [inset in Fig.~\ref{fig3}] is expected for $T<\Delta$, where $\Delta$ denotes the splitting of the lowest excited state from the GS doublet. It is due to crystal-field transitions driven by a magneto-elastic coupling \cite{Yaouanc}.
This single-ion mechanism leads to vanishing relaxation in the $T\rightarrow 0$ limit. In contrast, we observe saturation of the muon relaxation [Fig.~\ref{fig3}(b)], which, quite interestingly, occurs in SGS below the same temperature $T_c\sim 10$~K as previously found in NGS \cite{ZorkoNd}. The relaxation plateau evidences quantum spin fluctuations beyond the classical picture \cite{Simonet} setting up in both compounds and was observed before in several rare-earth-based frustrated magnets \cite{Dunsiger, Gardner, Dalmas, Bert}.

The entire $T$-dependence of $\lambda$ in both NGS and SGS can be fitted to the phenomenological model \cite{Salman,LancasterPRB}
\begin{equation}
\lambda(T) = \frac{1}{\frac{1}{\lambda_m}+A{\rm e}^{-\Delta/T}},
\label{eq4}
\end{equation}
valid for fast fluctuations when $\lambda \propto 1/\nu$.
In this model both the $T$-dependent phonon-induced broadening $A{\rm e}^{-\Delta/T}$ as well as the $T$-independent quantum contribution $1/\lambda_m$ are incorporated in the spectral width of spin fluctuations.  Our fits [Fig.~\ref{fig3}(b)] yield the low-$T$ relaxation rates $\lambda_m^{\rm NGS}=10.7$~$\mu$s$^{-1}$, $\lambda_m^{\rm SGS}=2.5$~$\mu$s$^{-1}$ and activation gaps $\Delta^{\rm NGS}=120$~K, $\Delta^{\rm SGS}=210$~K for NGS and SGS, respectively. The dynamical energy barrier $\Delta^{\rm NGS}$ agrees with the NSE experiment and is consistent with the static crystal-field splitting (90~K) between the GS and the lowest-lying crystal-field doublet \cite{Simonet}. We find that in SGS this splitting is even larger.

Next, we elucidate the magnetic GS of NGS and SGS and highlight the mechanism responsible for quantum relaxation leading to the relaxation plateau. Our density-functional-theory (DFT) calculations predict interstitial sites $P_1=(0.28, 0.22, -0.08)$ and $P_2=(0.27, 0.68,0)$ as the most probable muon stopping sites \cite{sup}, since electrostatic minima are found there (Fig.~\ref{fig1}). An estimate of the fluctuation rate $\nu = 2\gamma_\mu^2B_\perp^2/\lambda_m$ [Eq.~(\ref{eq2})] can be obtained by calculating the ensemble-averaged local magnetic field $B_{\perp}={\langle B_{\perp }^2\rangle}^{1/2}$. Since Langasites are insulators with well localized 4$f$ electrons, the dipolar contribution dominates. For NGS our calculations, taking into account noncorrelated moments $\mu_g^{\rm NGS}=1.6\mu_B$ within a sphere with a radius large enough to ensure convergence, yield $B_{\perp_1}=0.11$~T and $B_{\perp_2}=0.19$~T for $P_1$ and $P_2$, respectively. With $\lambda_m^{\rm NGS}=10.7$~$\mu$s$^{-1}$ these fields give on average $\nu^{\rm NGS}= 3.1$~GHz, a similar value as in our previous estimate for oxygen sites \cite{ZorkoNd}. This is in excellent agreement with the NSE experiment, which determined the spin correlation time $\tau^{\rm NGS}=1/\nu^{\rm NGS}\approx 0.3$~ns. Besides putting confidence in our DFT calculations, this accordance of complementary techniques confirms that muons directly probe the relaxation of electronic moments in Langasites and rules out any possible muon induced effect. The observation of the $\mu$SR relaxation plateau, therefore, proves intrinsic persistent spin dynamics in Langasites. The same analysis in SGS, taking into account the above determined $\mu_g^{\rm SGS}=0.25(3)\mu_B$ and $\lambda_m^{\rm SGS}=2.5$~$\mu$s$^{-1}$, yields the fluctuation rate $\nu^{\rm SGS}= 0.36(5)$~GHz, which is suppressed compared to $\nu^{\rm NGS}$ by $\sim\mu_g^{\rm NGS}/\mu_g^{\rm SGS}$.

\begin{figure}[t]
\includegraphics[trim = 10mm 4mm 1mm 6mm, clip, width=1\linewidth]{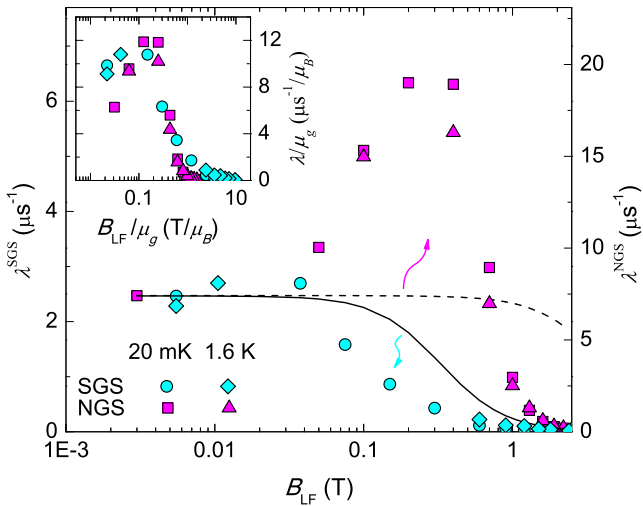}
\caption{Field dependence of the muon relaxation rate $\lambda$ in \SGS~(SGS) and \NGS~(NGS). Lines correspond to the model given by Eq.~(\ref{eq2}), with ZF parameters for NGS (dashed line) and SGS (solid line). Inset shows the universal scaling of $\lambda/\mu_g$ with $B_{\rm LF}/\mu_g$.}
\label{fig4}
\end{figure}

Our previous $\mu$SR investigation of NGS witnessed a pronounced field-induced maximum of $\lambda$ at 60 mK \cite{ZorkoNd}, contradicting Eq.~(\ref{eq2}) that would predict a monotonic decrease of $\lambda$ if the fluctuating rate $\nu$ was constant. We find that the maximum at $B_c^{\rm NGS}=0.3$~T is $T$-independent throughout the plateau region (Fig.~\ref{fig4}). Therefore, field polarization cannot play any role in its formation. Since $\gamma_\mu B_c^{\rm NGS}\ll \nu(B_{\rm LF}=0)$ the initial increase of $\lambda$ with increasing $B_{\rm LF}$ reveals decreasing fluctuating rate that reduces by a factor of $\sim 2$ at $B_c^{\rm NGS}$. This observation contradicts the traditional paramagnetic picture, in which a monotonic increase of $\nu$ due to a reduced energy barrier between the ground and excited states in the applied field \cite{MN} would monotonically decrease $\lambda$.
Similar maxima in $\lambda$ have been predicted \cite{LancasterJPCM2004} and observed \cite{LancasterJPCM2011} in MNMs and should be explained within the energy-level scheme of the system.

$B_c^{\rm NGS}$ sets a relevant energy scale in NGS, $\mathcal{J}^{\rm NGS}=\mu^{\rm NGS}_g B_c^{\rm NGS}/k_B\approx300$~mK, which nicely corroborates with inelastic neutron scattering \cite{ZhouNd} and specific-heat measurements \cite{Simonet} both suggesting ZF splitting of the GS Kramers doublet by $\sim$250~mK. This suggests that $\mathcal{J}^{\rm NGS}$ is associated with magnetic anisotropy, because the isotropic exchange cannot split the Kramers doublet. Such magnetic anisotropy cannot be ascribed to the dipolar interaction that is too small, $\mathcal{J}^{\rm NGS}_d=\mu^{\rm NGS}_g B_{d}/k_B\approx 25$~mK (average dipolar field is $B_{d} =25$~mT). This is further confirmed by the dipolar contribution to the Curie-Weiss temperature \cite{Daniels}
$\theta_d = \frac{\mu_0(\mu_g^{\rm NGS})^2}{4\pi k_B}\frac{1-\left(\mu_\perp^{\rm NGS}/\mu_g^{\rm NGS}\right)^4}{1+2\left(\mu_\perp^{\rm NGS}/\mu_g^{\rm NGS}\right)^2}
\sum_{j\ne i}\frac{3z_{ij}^2-r_{ij}^2}{r_{ij}^5}=-13$~mK that is an order of magnitude below the experimental value $\theta=-120(20)$~mK. Thus, we attribute the energy scale $\mathcal{J}^{\rm NGS}$ to the anisotropic part of the tensorial exchange interaction, which is for rare earths with large spin-orbit coupling often of the same magnitude as the isotropic part \cite{Thompson}. We note that $\theta$ includes contributions of both parts and is, therefore, also in agreement with $\mathcal{J}^{\rm NGS}$.

In SGS we find a similar field dependence of the relaxation rate, provided that both $B_{\rm LF}$ and $\lambda$ are normalized by the moment value $\mu_g$ (inset in Fig.~\ref{fig4}). This evidences the characteristic scale $B_c / \mu_g \sim 0.1$~T/$\mu_B$, independent of the rare-earth ion. It yields  $\mathcal{J}^{\rm SGS}/\mathcal{J}^{\rm NGS}\sim \left( \mu_g^{\rm SGS}/\mu_g^{\rm NGS} \right)^2 $, further supporting our claim that the energy scale $\mathcal{J}$ should be attributed to a pairwise interaction. In this case, the expected dependence $\nu/\mu_g= f(B_{\rm LF}/\mu_g)$ according to Eq.~(\ref{eq2}) leads to the scaling $\lambda/\mu_g\propto\frac{f(B_{\rm LF}/\mu_g)}{f^2(B_{\rm LF}/\mu_g)+(B_{\rm LF}/\mu_g)^2}$ . The pairwise interaction $\mathcal{J}$ should bring about building up of spin correlation for $T\lesssim\mathcal{J}$. This could indeed be responsible for the prominent decrease of $\lambda^{\rm NGS}$ below $\sim$0.5~K [Fig.~\ref{fig3}(b)], which is absent in SGS in the same temperature range because the relevant energy scale $\mathcal{J}^{\rm SGS}$ is much smaller.

Persistent fluctuations at temperatures $T/\Delta\sim 10^{-4}$ in NGS and SGS can be explained by QT through a double-well potential separating the two states of the crystal-field Kramers doublet. A matrix element that couples these two states and induces tunneling has to be associated with magnetic anisotropy, resulting in a splitting of the GS. In MNMs various sources of anisotropy have been invoked, such as transverse single-ion anisotropy, hyperfine interaction, dipolar fields and anisotropic exchange interactions \cite{MN}. Since in NGS and SGS the muon-relaxation plateau persists to temperatures far above $\mathcal{J}$, single-ion spins are tunneling due to quasi-static local magnetic fields that break the time-reversal symmetry. The single-ion anisotropy that leads to QT in integer-spin MNMs \cite{MN} is ineffective in half-integer-spin Langasites due to topological reasons \cite{Loss,Wernsdorfer}. In half-integer-spin MNMs quasi-static nuclear hyperfine fields have proven responsible for muon relaxation \cite{LancasterPRB,Keren}. However, these cannot explain QT in Langasites, because the hyperfine interaction in Nd$^{3+}$ and Sm$^{3+}$ is of the same magnitude, $A_J\sim10$~mK \cite{AB}, which should lead to very similar fluctuation rates $\nu^{\rm NGS}$ and $\nu^{\rm SGS}$; e.g., as found for a family of isotropic MNMs \cite{Keren}. Moreover, in both materials the natural abundance of magnetic rare-earth nuclei is below 30\%, 
thus the stretch exponent $\alpha$ would be different inside and outside of the QT regime. Finally, since the dipolar interaction is much smaller than $\mathcal{J}$, we assign the anisotropic exchange interaction between rare-earth moments as being responsible for QT in Langasites. It produces longitudinal as well as perpendicular effective magnetic fields, the latter inflicting QT \cite{Prokofev}.

We propose a picture of tunneling under "quenched" anisotropic-exchange fields, where stochastic fluctuations of the environment are slow compared to spin correlation times $\tau^{\rm NGS}= 0.33$~ns and $\tau^{\rm SGS}= 2.7$~ns of reference spins. This condition is satisfied when the thermally-assisted spin relaxation becomes slower than the quantum relaxation [Eq.~(\ref{eq4})], which occurs around 20~K in both systems and thus rationalizes the materialization of the quantum relaxation at temperatures two orders of magnitude above the dominant interaction $\mathcal{J}$.

In conclusion, we suggest that the persistent spin dynamics in the Kramers representatives of the kagome-lattice Langasites are due to QT and are triggered by short-ranged magnetic anisotropy splitting the GS doublet. Similarly, a tunneling process between different equivalent spin configurations has been proposed to explain the dynamics of spin-ice pyrochlores above freezing \cite{Ehlers,Snyder}.
QT favors a superposition of degenerate states over selecting a particular state. Therefore, it may affect frustrated lattices profoundly, possibly leading to a quantum spin-liquid GS \cite{Shannon},
 like the dynamical quantum spin-ice state \cite{Onoda}  expected to materialize in several pyrochlores \cite{Onoda, Molovian, Zhou} with substantial anisotropic exchange interaction.
QT thus appears as a momentous feature of the geometrically frustrated rare-earth-based compounds and a close resemblance exists between these magnetically anisotropic systems and mesoscopic molecular nanomagnets.

\acknowledgments
We thank A.~Keren, S.~J.~Blundell and T.~Lancaster for stimulating discussions, and C.~Paulsen for allowing us to use his SQUID dilution magnetometer. This work was partly supported by the ANR-09-JCJC-0093-01 grant. AZ acknowledges financial support of the Slovenian Research Agency (Projects J1-2118, BI-FR/11-12-PROTEUS-008).

\appendix

\begin{widetext}
\vspace{19cm}
\begin{center}
{\large {\bf Supplementary information:\\

Quantum Tunneling in Half-Integer-Spin Kagome-Lattice Langasites}}\\
\vspace{0.5cm}
A. Zorko,$^{1,2}$ F. Bert,$^3$ P. Mendels,$^3$ A. Poto\v{c}nik,$^1$ A. Amato,$^5$ C. Baines,$^5$\\
K. Marty,$^6$ P. Bordet,$^7$ P. Lejay,$^7$ E. Lhotel,$^7$ V. Simonet,$^7$ and R. Ballou$^7$
\vspace{0.2cm}

{\it
$^1$Jo\v{z}ef Stefan Institute, Jamova 39, SI-1000 Ljubljana, Slovenia\\
$^2$EN--FIST Centre of Excellence, Dunajska 156, SI-1000 Ljubljana, Slovenia\\
$^3$Laboratoire de Physique des Solides, Universit\'e Paris-Sud 11, UMR CNRS 8502, 91405 Orsay, France\\
$^4$Institut Universitaire de France, 103 Boulevard Saint-Michel, F-75005 Paris, France\\
$^5$Laboratory for Muon Spin Spectroscopy, Paul Scherrer Institut, CH-5232 Villigen PSI, Switzerland\\
$^6$SPINTEC, (UMR8191 CEA/CNRS/UJF/G-INP), CEA Grenoble, INAC, 38054 Grenoble, France\\
$^7$Institut N\'eel, CNRS and Universit\'e Joseph Fourier, BP 166, 38042 Grenoble, France\\}
\end{center}
\end{widetext}

\section {Synthesis and Crystal Structure}

\begin{figure}[t]
\includegraphics[width=1\linewidth, trim=0 0 0 0, clip=true]{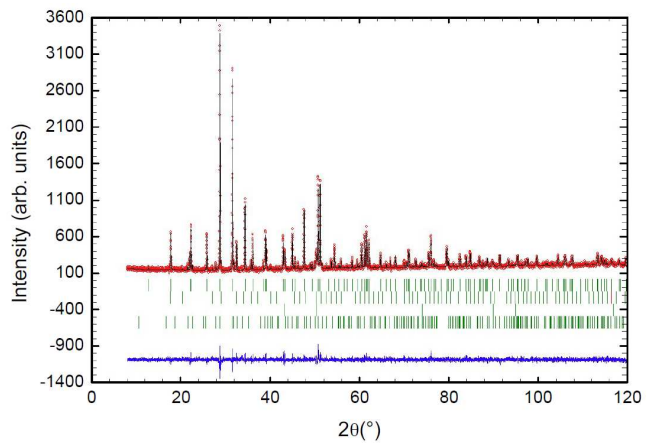}
\caption{SGS x-ray diffraction pattern at room temperature (red dots) with its Rietveld refinement (black line) showing the presence of 88.7(6) mass \% of the Sm$_3$Ga$_{2.63}$Al$_{2.37}$SiO$_{14}$ Langasite, 5.8(1) mass \% of the Sm$_3$Ga$_5$O$_{12}$ garnet and 5.5(1) mass \% of Sm$_{4.66}$O(SiO$_4$)$_3$ \cite{Hartenbach}, as well as some contamination of the Cu sample holder. The difference between the measurement and the calculation is shown as a blue line.}
\label{fig1s}
\end{figure}

Powders of NGS and SGS compounds were prepared by solid
state reactions of stoichiometric amounts of high purity oxides at
1420$^{\circ}$C, resp. 1400$^{\circ}$C in air. X-ray powder diffraction patterns
were measured to confirm the crystal structures reported in Ref.
\cite{Iwataki}. In the SGS Langasite with a heavier rare earth atom, part of Ga has to be replaced by the smaller Al in order to stabilize the structure. 
The Rietveld refined diffraction pattern of SGS is shown in Fig.~\ref{fig1s} and the structural parameters are reported in Tab.~\ref{struc}. Except from the Sm site, all other cation sites show substitutions. The Si cation occupancy is fixed by charge balance, and the refinement of the Ga/Al mixed occupancies led to the overall composition for SGS; Sm$_{3}$Ga$_{2.63}$Al$_{2.37}$SiO$_{14}$. The NGS single crystal was grown by the floating zone method using an image furnace, under a 99\% Ar + 1\% O$_2$ atmosphere, at a growth rate of 10 mm/h. 

\begin{table*}[t]
\caption{Structural parameters of SGS determined by x-ray powder diffraction with a D8 Brucker diffractometer. The data refined with the Rietveld method using the Fullprof software \cite{Fullprof} led to
Bragg R-factor= 10.9, Rf-factor= 8.35, $\chi^2$=1.4 and the cell parameters a = 7.94798 (3) {\AA}, c = 4.96578 (3) {\AA} at room temperature. The Ga3/Si3 site is half occupied by Si and a mixture of Ga/Al. The columns are the atom type, the Wyckoff site, the atomic positions, the isotropic displacement parameter and the occupancy.}
\begin{center}
\begin{ruledtabular}
\begin{tabular}
{c c c c c c c} Atom & Site & X & Y & Z & B$_{iso}$ & Occ. \\\hline Sm & 3e & 0.4157(2) & 0 & 0 & 0.5 & 1 \\ Ga1 & 1a & 0 & 0 & 0 & 1.074(7) & 0.432 \\ Al1 & 1a & 0 & 0 & 0 & 1.074(7) & 0.564 \\ Ga2 & 3f & 0.7706(9) & 0 & 1/2 & 1.074(7) & 0.536 \\ Al2 & 3f & 0.7706(9) & 0 & 1/2 & 1.074(7) & 0.464 \\ Ga3 & 2d & 1/3 & 2/3 & 0.4603(31)& 1.074(7) & 0.294 \\ Al3 & 2d & 1/3 & 2/3 & 0.4603(31) & 1.074(7) & 0.204\\ Si3 & 2d & 1/3 & 2/3 & 0.4603(31) & 1.074(7) & 0.501 \\ O1 & 2d & 1/3 & 2/3 & 0.8025(54) & 1.0 & 1 \\ O2 & 6g & 0.4638(40) & 0.3169(23) & 0.7032(28) & 1.0 & 1 \\ O3 & 6g & 0.2301(22) & 0.0783(18) & 0.2514(22) & 1.0 & 1
\end{tabular}
\end{ruledtabular}
\end{center}
\label{struc}
\end{table*}

\section {Determination of Muon Stopping Sites}

Positively charged muons favor electronegative environment. In oxydes, it is generally believed that a muon stops close to oxygen, at a distance of about 1~\AA~\cite{Yaouanc}. In order to determine possible muon stopping sites in Langasites we employed density-functional-theory (DFT) calculations for the isostructural La$_3$Ga$_5$SiO$_{14}$ compound. The goal was to determine a self-consistent electron-density distribution, which yielded a spatial profile of the electrostatic potential. Similar calculations have proven successful in the past \cite{Luetkens}.

Our calculations were performed using the pwscf program, a part of the Quantum Espresso software package \cite{Giannozzi}. Ultrasoft pseudo potentials appropriate for the Perdew-Burke-Erzerhof exchange-correlation (LDA) were used. The pseudo Bloch functions were expanded over plane waves with the energy cutoff of 35~Ry on a $6\times6\times6$ Monkhorst-Pack \textit{k}-space mesh. The convergence of the results and the total energy was tested and yielded negligible variation at this values. The Fermi surface was smeared with a ``temperature" parameter of 2~meV. The computed electrostatic potential consisted of bare ionic and Hartree electrostatic potentials.

A global electrostatic potential minimum was found at the interstitial site $P_1=(0.28, 0.22, -0.08)$ with multiplicity of six (6g), while local electrostatic minima were found at $P_2=(0.27, 0.68,0)$ and $P_{2'}=(0.59, 0.32, 0)$, both with multiplicity of three (see Fig.~1 of the main part). The value of the potential at $P_2$ is only slightly enhanced with respect to the global minimum. Different value of the potential at crystallographically equivalent $P_2$ and $P_{2'}$ is due to their proximity to the Ga(3) crystallographic site, which is randomly occupied with Ga$^{3+}$ and Si$^{4+}$ \cite{Bordet}. In the case of Si$^{4+}$ the local minimum becomes significantly less intense. As expected, all electrostatic minima are located in a vicinity of the most electronegative O$^{2-}$ ions, $P_1$ at the distance of 1.27~\AA~and $P_2$ ($P_{2'}$) 0.93~\AA~away, in agreement with the generally accepted distance \cite{Yaouanc}. Assuming that the rather localized 4$f$ electrons do not appreciably affect the electrostatic picture except in the close vicinity of rare earths and that the muons do not perturb the crystal structure, we assign $P_1$ and $P_2$ as the most probable muon stopping sites also in \NGS~and \SGS. The same stretched-exponential functional form of the muon depolarization curves ($\alpha=0.6$; see main part) in both materials suggests that small structural differences between them have no major impact on the muon stopping sites.

\end{document}